\documentclass[10pt,letterpaper]{article}
\usepackage[top=0.85in,left=2.75in,footskip=0.75in]{geometry}
\usepackage{amsmath,amssymb}
\usepackage{changepage}
\usepackage[utf8x]{inputenc}
\usepackage{textcomp,marvosym}
\usepackage{cite}
\usepackage{nameref,hyperref}
\usepackage[right]{lineno}
\usepackage{microtype}
\DisableLigatures[f]{encoding = *, family = * }
\usepackage[table]{xcolor}
\usepackage{array}
\usepackage[normalem]{ulem}
\newcolumntype{+}{!{\vrule width 2pt}}

\newlength\savedwidth

\raggedright
\usepackage[aboveskip=1pt,labelfont=bf,labelsep=period,justification=raggedright,singlelinecheck=off]{caption}

\makeatletter
\renewcommand{\@biblabel}[1]{\quad#1.}
\makeatother

\usepackage{lastpage,fancyhdr,graphicx}
\usepackage{epstopdf}
\fancyheadoffset[L]{2.25in}
\fancyfootoffset[L]{2.25in}
\begin{document}
\vspace*{0.2in}

\begin{flushleft}
{\Large
\textbf
\newline
The role of dose-density in combination cancer chemotherapy
}
\newline

\'{A}lvaro G. L\'{o}pez\textsuperscript{1*},
Kelly C. Iarosz\textsuperscript{2},
Antonio M. Batista\textsuperscript{3},
Jes\'{u}s M. Seoane\textsuperscript{1},
Ricardo L. Viana\textsuperscript{4},
Miguel A. F. Sanju\'{a}n\textsuperscript{1}\\
\bigskip
\textbf{1} Departamento de F\'isica, Universidad Rey Juan Carlos, M\'ostoles, Madrid, Spain \\
\textbf{2} Institute of Physics, University of S\~{a}o Paulo, S\~{a}o Paulo, SP, Brazil \\
\textbf{3} Department of Mathematics and Statistics, State University of Ponta Grossa, Ponta Grossa, PR, Brazil\\
\textbf{4} Department of Physics, Federal University of Paran\'{a}, Curitiba, PR, Brazil\\
\bigskip

*alvaro.lopez@urjc.es

\end{flushleft}

\section*{Abstract}
A multicompartment mathematical model is presented with the goal of studying the role of dose-dense protocols in the context of combination cancer chemotherapy. Dose-dense protocols aim at reducing the period between courses of chemotherapy from three to two weeks or less, in order to avoid the regrowth of the tumor during the meantime and achieve maximum cell kill at the end of the treatment. Inspired by clinical trials, we carry out a randomized computational study to systematically compare a variety of protocols using two drugs of different specificity. Our results suggest that cycle specific drugs can be administered at low doses between courses of treatment to arrest the relapse of the tumor. This might be a better strategy than reducing the period between cycles.

\section*{Author summary}

Chemotherapeutic drugs for the treatment of cancer are currently delivered in periodic cycles to allow healthy tissues to recover from their cytotoxic effects. Unfortunately, this recovery window allows the tumor to regrow between cycles as well. It has been observed that a reduction of the periodicity from three to two weeks can be beneficial without introducing higher toxicities. These new chemotherapeutic protocols are known under the term dose-dense chemotherapy. We perform a computational study to assess the limitations of the dose-dense approach and suggest possible alternatives.


\section*{Introduction}\label{sec:intro}

Overcoming the toxic side-effects of chemotherapeutic drugs on healthy tissues in the short term, as well as the development of drug resistance of cancer cells to such medications in the long run, are the two key cornerstones in the improvement of cytotoxic drug therapy \cite{perry,lipo,drr}. These two limiting factors put before, the rest of our research efforts must concentrate on the particular arrangement of protocols. Randomized trials carried out along the last forty years have demonstrated that the way in which protocols are organized is of great relevance as well \cite{dd1,dd2,dd3,clinclast}. For instance, breast cancer trialists have revealed that, in addition to the drug dose-intensity, the periodicity of the cycles of chemotherapy affects the patients' outcome as well \cite{dd4}. Accordingly, a sequential administration of two drugs can be more beneficial than the same drugs given alternately  \cite{dd1}. In particular, a reduction of the period between cycles of chemotherapy from three to two weeks has introduced moderate but statistically significant benefits in disease free survival at five years and overall survival \cite{dd4}. Fortunately, this can be achieved without entailing higher toxicities with the assistance of granulocyte colony-stimulating factor. 

The increase in the frequency of drug administration has been termed dose-dense chemotherapy, because when the dose is graphically represented against time, these protocols look more tight \cite{norton}. Of course, from a physical point of view, the temporal density of a certain magnitude can also be raised by increasing the amount of such physical magnitude while keeping the same distribution over time. On the other hand, dose-intensity is commonly defined in the field of oncology as an average by dividing the total dose of a drug administrated during a treatment by the duration of such treatment \cite{dd4}. Thus, as currently defined and utilized, dose-density and dose-intensity are two mudding and redundantly interrelated concepts. To avoid any sort of confusion we unwind these two ideas by adopting the following criteria \cite{ddpc}. The term dose-density is here reserved to denote a variation in the periodicity of the cycles of chemotherapy when the dose per cycle is kept constant, while the term dose-intensity of a drug is hereinafter referred to an increase in the amount of dose administered per cycle.

From a theoretical point of view, the ground on which dose-dense chemotherapy settles is known as the Norton-Simon hypothesis. This hypothesis states that the rate of destruction by chemotherapeutic drugs is proportional to the rate of growth of an identical unperturbed tumor \cite{nsh}. This statement is founded on the notion of mitotoxicity, which proclaims that chemotherapy acts more efficiently on the cells that are rapidly proliferating within a tissue \cite{perry}. Consequently, since only a fraction of the cells that shape a solid tumor are actively dividing \cite{capas,bru}, and bearing in mind that more cells in proportion turn into the quiescent state so as to render the growth of the tumor sigmoid \cite{gomp}, the impact of chemotherapy should be stronger for smaller tumors. This effect would hinder the efficiency of chemotherapy because, at some point of the growth curve, the regrowth between cycles would prevail over the destruction caused by the drugs during that same cycle. Note that tumors tend to regrow more rapidly after a cycle of chemotherapy, becoming less sensitive at the end of such period of time, when the subsequent cycle arrives.

In summary, the progressive doubling time dilation of a growing solid tumor defies the log-kill paradigm that has dominated traditional chemotherapy over decades \cite{lokil}, which is being curtailed to non-solid malignancies nowadays. These supporting arguments notwithstanding, it has been recently demonstrated that dose-density might be pertinent even in the absence of any hypothesis affecting the particular nature of the tumor's growth \cite{ddpc}. However, enthusiasm must be prevented at the same time, since the benefits of dose-dense chemotherapy are still being the subject of intense experimental research and scientific debate \cite{notwo}. We recall that the biological mechanism by which a certain drug inflicts its damage in practice might not correspond exactly to the one for which it was originally conceived \cite{prp}.

Bearing all these facts in mind, the appearance of dose-dense protocols for the treatment of solid tumors has been at all events an important breakthrough in breast and ovarian cancer. Furthermore, it is also posing crucial questions concerning the nature of solid tumor growth and its dynamical response driven by complex combination therapies \cite{norton}. Always bounded to experimental verification, needless to say, many of these questions can be better addressed within the framework of mathematical and \emph{in silico} modeling \cite{founcanmod}. In the present work we propose a multidimensional ordinary differential equation (ODE) model for cancer chemotherapy to investigate under which conditions dose-dense protocols are more or less advantageous. As far as we are concerned, the present work represents the first mathematical and computational approach to the study of the dose-dense principle in a cytotoxic combination therapy \cite{train,castor,monro,dono,cold}. We disregard toxicity models in this study, since it is not our purpose to develop specific protocols for a certain class of tumors, which could sometimes be regarded with skepticism \cite{putten}. We would rather find general guiding principles that may aid clinicians to bias their decisions in their proposal of better forthcoming chemotherapeutic protocols.

\section{Methods} \label{sec:M}

\subsection{Model description}\label{sec:md}
\begin{figure}
\centering
\includegraphics[width=0.7\linewidth,height=0.3\linewidth]{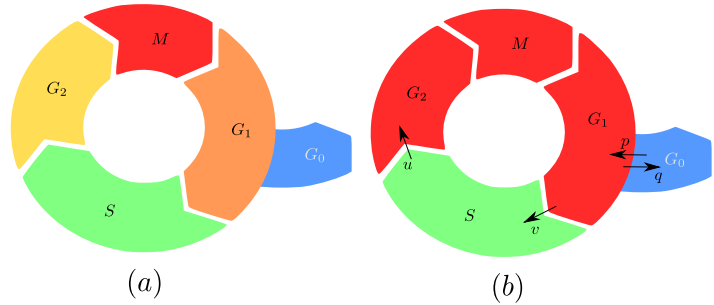}
\caption{\textbf{Schematic representation of the cell cycle}. (a) In the life of an eukaryotic cell, a new individual is born by mitotis ($M$) and starts its journey at Gap 1 ($G_{1}$). If the conditions are adequate, it proceeds to replicate its DNA during the phase of synthesis ($S$). If they are not, the cell exists the cycle and becomes quiescent ($G_{0}$). After DNA synthesis the cell resumes its growth through the Gap 2 phase ($G_{2}$) and, if there is no DNA damage, a new mitosis begins. The whole cycle lasts around twenty four hours for a typical proliferating human cell \cite{cohau}. (b) Three phases of the cell cycle are joined into a single compartment, yielding a new model where the arrows represent the flow direction and constant rates between compartments.}
\label{fig:1}
\end{figure}

A general multicompartment model for cancer chemotherapy can be built from preceding modeling attempts \cite{panad,panet, pinho, gard2,kav,kavg}. Since the cell cycle comprises four phases (mitosis, Gap 1, synthesis and Gap 2) and cells can also be quiescent (Gap 0), we consider a model with five different compartments representing the cell populations in each of them (see Fig.~\ref{fig:1}(a)).

The use of different compartments is relevant since some drugs destroy only cells in a specific phase of the cell cycle. We assign to each of these compartments a number $j$ between one and five, respectively. Therefore, the five cell populations can be represented by a vector $\vec{P}(t)$, whose components can be written in tensor form as $P_{j}(t)$. We allocate the first compartment $(j=1)$ to mitotic cells, while quiescent cells are inserted in the last compartment $(j=5)$. Following previous works \cite{gard2}, we assume that only quiescent cells die from necrosis. On the contrary, the apoptotic pathway might be activated in any compartment \cite{apop}. However, tumors frequently evolve towards states of low apoptotic index quickly \cite{gard2}. Consequently, and for computational simplicity, we sweep apoptosis under the rug of the proliferation term, by considering that this pathway simply reduces the net growth rate of tumor cells. Finally, every equation is equipped with two more terms. The first of them represents the rate of flow between adjacent compartments, while the second one models the action of cytotoxic drugs on each compartment. Broadly speaking, cytotoxic drugs can be classified as cycle specific (CS) or cycle non-specific (CNS). The former inflict their damage at a specific point of the cycle of the cell, while the second can destroy cells independently of their kinetic status. Therefore, the two vector variables $\vec{C}_{s}(t)$ and $\vec{C}_{n}(t)$ symbolize the different types of drugs that have to be delivered to the patient, respectively. The dimension of each vector is given by the number of drugs scheduled in the treatment. In summary, the differential equation governing the dynamics of each population $P_{j}(t)$ in the system can be written as
\begin{equation}
\dfrac{d P_{j}}{d t}= \delta_{1 j} g(\vec{P}) P_{j}-\delta_{5 j}n(\vec{P})P_{j} +\sum^{5}_{i=1}w_{j i}(\vec{P})P_{i}-k_{j}(\vec{P},\vec{C}_{s},\vec{C}_{n}) P_{j},
\label{eq:1}
\end{equation}
where $\delta_{i j}$ is the Kronecker delta and $j=1,...,5$. As thoroughly described ahead, the functions $g(\vec{P})$ and $n(\vec{P})$ are nonlinear functions representing the fractional cell growth of the tumor cells in the mitotic compartment and the fractional cell death due to necrosis in the quiescent compartment. The functions $w_{j i}(\vec{P})$ and $k_{j}(\vec{P},\vec{C}_{s},\vec{C}_{n})$ are nonlinear functions representing the fractional cell flow between compartments and the fractional cell kill of chemotherapeutic drugs on each compartment, respectively. If the cell cycle is regarded as a Markov chain, the terms $w_{j i}(\vec{P})$ are related to the transition matrix. On the other hand, the detailed functions $\vec{C}_{s}$ and $\vec{C}_{n}$ depend on the pharmacokinetical model used and the way in which drugs are administered. Note that this inhomogeneous five-dimensional model is constituted by two well differentiated parts. The three first terms on the right hand side of equation~\eqref{eq:1} define the cell kinetics and include the turnover of cells from one stage of the cell cycle to another, as well as the birth of tumor cells and their death from other causes different from cytotoxic treatment. The remaining fourth term on the right hand side of the same equation represents precisely the destruction of tumor cells by the cytotoxic agents. In the next lines we particularize the present model to meet the needs of the present work.
\begin{figure}
\centering
\includegraphics[width=0.48\linewidth,height=0.44\linewidth]{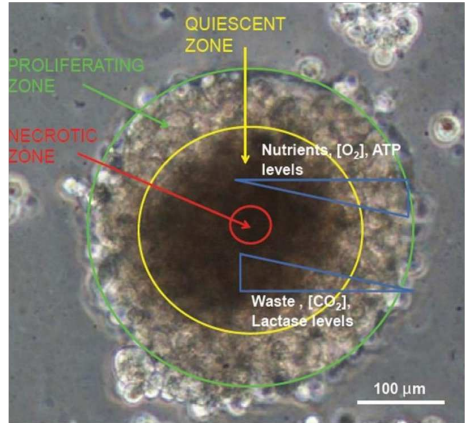}
\caption{\textbf{A tumor spheroid}. A spherical tumor showing several layers  with cell populations displaying different kinetic status. The various gradients of chemicals from the interior to the outermost layer are schematically represented. As can be seen, only a fraction of cells, placed at the most exterior region, are actively proliferating. The different layers correspond to different compartments of the model (from Ref.~\cite{capas} under license CC BY-SA 4.0).}
\label{fig:2}
\end{figure}

\subsection{Kinetic model}
In practice, one or more compartments of the previous general model can be merged into a single compartment, reducing the dimensionality of the problem (see Fig.~\ref{fig:1}(b)). For example, in the present work, there is only one CS drug and another CNS drug. Therefore, only three kinetic compartments are needed. If we assume that the cycle specific drug only targets cells in the phase of synthesis, as it frequently occurs, two compartments are required to represent cells that are actively dividing. One of these compartments contains cells that are in the synthesis phase at a particular instant of time $t$. This cell population is represented henceforth by the variable $S(t)$, while the remaining cells comprise all those cells that are in the complementary part of the cell cycle (mitosis, Gap 1 and Gap 2). We just gather these three compartments into a single one and refer to the cell population contained in it as mitotic, representing it mathematically by the letter $M(t)$. To conclude, one more cell population $Q(t)$ is included to model those resting cells that make up the quiescent compartment (see Fig.~\ref{fig:2}). Therefore, in this first study we shall utilize the three-dimensional model given by the following system of differential equations
\begin{equation}
\begin{array}{lr}
\dfrac{d M}{d t}=(r-v)M+u S-w(P) M+p Q \bigskip\\
\dfrac{d S}{d t}=-u S+v M \bigskip\\
\dfrac{d Q}{d t}=w(P) M-(p+n)Q,
\end{array}
\label{eq:2}
\end{equation}
where $r$ stands for the net growth constant rate resulting from proliferation and apoptosis of the cells in the mitotic pocket and $n$ describes the rate at which quiescent cells die from necrosis. These terms correspond to the functions $g(\vec{P})$ and $n(\vec{P})$ in equation~\eqref{eq:1}, which are chosen to be constants. The parameters $u,v$ are the flow constant rates of cells from the s-phase to the mitotic population and vice versa. The term $w(P)$ and $p$ are the rates describing the speed at which cells flow from the mitotic compartment to the $G_{0}$ phase and vice versa. Now, these functions correspond to the functions $w_{ji}(\vec{P})$ in the original model. The variable $P(t)$ represents the whole tumor cell population resulting from adding the cells in all the compartments, \emph{i. e.}, the relation $P(t)= M(t)+S(t)+Q(t)$ holds. Note that the total tumor cell population is governed by the net growth rate balanced by the rate of necrosis
\begin{equation}
\begin{array}{lr}
\dfrac{d P}{d t}=r M-n Q.\\
\end{array}
\label{eq:3}
\end{equation}

Most of the fractional cell functions are considered constant for simplicity, following previous works \cite{gard2} and traditional quasispecies models \cite{quasispecies}. However, the flow rate between mitotic cells and quiescent cells is more generally assumed to be governed by the linear term $w(P)M$. The reason why this is so is that we want the model to represent limited growth following a sigmoid function up to a certain maximum number of tumor cells $P^{*}= K$. This stable fixed value is frequently known as the carrying capacity of the tumor and we shall follow this terminology here. More specifically, we assume logistic growth \cite{reacdis,mixed,validbul} for mathematical simplicity.  Nevertheless, other types of sigmoid functions, such as Gompertzian growth or a Richards' type of growth can be used instead, since they are all generated by diffusive long range cell interactions \cite{mftcg}. In short, we consider that the flow rate of mitotic cells to the resting pocket obeys a linear function $w(P)=q P$, which gives rise to a logistic type of growth.

We now describe how to compute the model parameters describing the rates of flow $u,v,q$ between compartments in terms of the other parameters, which are more amenable to the experimental practice and for which we have several values in the literature \cite{gard2}. These better known parameters are the constant rate of mitosis $r$, the turnover of mitotic to quiescent cells $p$, the proliferative index, here denoted as $\wp$ and defined as
\begin{equation}
\wp(t)= \dfrac{M(t)+S(t)}{P(t)},
\label{eq:4}
\end{equation}
and the fraction of time $f$ that cells spend in the phase of synthesis.

The parameters $u$ and $v$ can be expressed in terms of $r$ by considering the time spent in each phase of the cycle by a human cell. These constant rates can be estimated by means of the following argument. If we name the total cell cycle time by $\tau$ and the cell cycle of the synthesis phase is denoted as $\tau_{S}$, while the remaining time of the cycle is written as $\tau_{M}$, we have $\tau= \tau_{S}+\tau_{M}$. In particular  $\tau_{S}=f\tau$ and $\tau_{M}=(1-f)\tau$ are fulfilled. Now, for a geometric growth, the doubling time is inversely proportional to $r$, which allows us to write $u=r/f$ and $v=r/(1-f)$. On the side of the variables we have that at equilibrium the rates are zero $d M/d t=d S/d t=d Q/d t=0$ and, consequently, also the relation $d P/d t=0$ holds. Now, we name the values of the cell populations at equilibrium as $M^{*}$, $S^{*}$ and $Q^{*}$ and assume that they are different from zero. Then, bearing in mind that $P^{*}=M^{*}+S^{*}+Q^{*}=K$, it is straightforward to obtain from the second identity appearing in equation~\eqref{eq:2} that $S^{*}=M^{*}v/u$, and from equation~\eqref{eq:3} that $Q^{*}=M^{*} r/n$. Using these two results and the equations obtained from the conditions $d M/d t=d S/d t=0$, the constraint
\begin{equation}
q=\frac{r}{K} \left(1+\frac{p}{n}\right)
\label{eq:5}
\end{equation}
is implied. Furthermore, making use of equation~\eqref{eq:2} we can also derive the size of each cell population at equilibrium, which reads
\begin{equation}
M^{*}=\dfrac{K}{1+v/u+r/n},~S^{*}=\dfrac{K v/u}{1+v/u+r/n}~\text{and}~Q^{*}=\dfrac{K r/n}{1+v/u+r/n}.
\label{eq:6}
\end{equation}
From these three values it is easy to check that the proliferative index at equilibrium is written as
\begin{equation}
\wp^{*}=\dfrac{1+v/u}{1+v/u+r/n},
\label{eq:7}
\end{equation}
from which we can derive the following parameter relation for the constant rate of necrosis $n$ as
\begin{equation}
n=r \left(1-f\right)\left(\dfrac{\wp^{*}}{1-\wp^{*}}\right).
\label{eq:8}
\end{equation}
In the numerical study conducted in the section containing our results we present all the parameter values of the model, which are derived from the elementary parameters $r,f,\wp^{*},p$ by means of the preceding equations.

\subsection{The fractional cell kill of cytotoxic drugs}

Once the kinetic model and its parameters have been established, we proceed to introduce the fractional cell kill of cytotoxic drugs. Just as a reminder, and because sometimes the term rate is loosely used to refer a constant rate \cite{validpilis,onlaw} or the fractional cell kill itself \cite{onlaw}, we clarify this concept. It describes the rate at which a cell population is reduced to a fraction of itself after a fixed period of time. If the cell population presents a constant fractional killing, it means that it decays exponentially in time. The fractional cell kill should be then identified with the constant rate of the decay, being both in direct relation. Therefore, the fractional cell kill must be rigorously defined from a mathematical point of view as the rate at which the logarithm of a cell population $P$ decreases as a consequence of some biological process ($d \ln P/d t$), as it is done in previous works \cite{onlaw}.

In the present case, such attrition is caused by cytotoxic agents. To work out the mathematical nature of the function $k(P,C)$ that describes the cell kill by a chemotherapeutic drug we appeal to past works on this subject \cite{ncc}. There, the following function is proposed
\begin{equation}
k(P,C)=\dfrac{\mu}{1+ \sigma P/K}\left(1-e^{-\rho C}\right),
\label{eq:9}
\end{equation}
where $C(t)$ is the concentration of the drug at the tumor site and $\mu$ is the maximum fractional cell kill achievable by the drug, when it is ideally given at an infinite dose to sufficiently small (exponentially growing) tumors. The term $1-e^{-\rho C}$ is based on the Exponential Kill Model \cite{gard}. This mechanistic model was developed in conformity with \emph{in vitro} data and has also been tested against \emph{in vivo} experiments with mice \cite{gard, validbul}. Contrary to traditional log-kill models, it exhibits saturation of the cell kill as the drug concentration is raised. The steepness of the dose-response curve is related to the sensitivity $\rho$ of a tumor cell line to a particular drug. Simply put, for a nascent tumor the parameter $\mu$ would be the maximum constant rate of destruction achievable at high doses. On the other hand, the parameter $\rho$ tells us how the response to the drug is modified as the cells become increasingly resistant. This reduction in sensitivity to the drugs is reflected by a smaller value of the parameter $\rho$. Finally, in another investigation \cite{ncc} we derive and explain in great detail the Michaelis-Menten dependence of the cell kill rate on the total tumor size population $P(t)$. It has been introduced in the model in order to account for the Norton-Simon hypothesis, whose intensity is condensed in the parameter $\sigma$. As the tumor gets bigger and approaches its carrying capacity, the maximum fractional cell kill reduces from $\mu$ to the value $\mu/(1+\sigma)$, with $\sigma>0$.
\begin{figure}
\centering
\includegraphics[width=0.7\linewidth,height=0.23\linewidth]{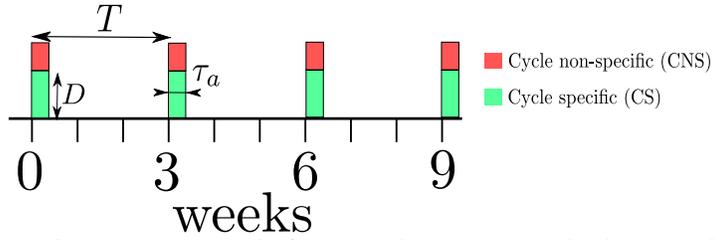}
\caption{\textbf{Chemotherapy protocol}. A protocol is represented schematically, showing the function $I(t)$ for each drug. In particular, the size of each bar represents the value of this function. In this figure we showed the CS and CNS drugs delivered concurrently with a period $T$ of three weeks using different doses $D$ for each of them. The time duration of each infusion $\tau_{a}$, however, is the same for the two drugs.}
\label{fig:3}
\end{figure}

Since we only deliver one cycle specific drug $C_{s}$ and another cycle non-specific drug $C_{n}$ in the foregoing investigation, the whole set of differential equations together with chemotherapy can be written as
\begin{equation}
\begin{array}{lr}
\dfrac{d M}{d t}=F_{M}(M,S,Q)-\dfrac{\mu_{n}}{1+ \sigma_{n} P/K}\left(1-e^{-\rho_{n} C_{n}}\right)M \bigskip\\
\dfrac{d S}{d t}=F_{S}(M,S,Q)-\dfrac{\mu_{n}}{1+ \sigma_{n} P/K}\left(1-e^{-\rho_{n} C_{n}}\right)S-\dfrac{\mu_{s}}{1+ \sigma_{s} P/K}\left(1-e^{-\rho_{s} C_{s}}\right)S \bigskip\\
\dfrac{d Q}{d t}=F_{Q}(M,S,Q)-\epsilon \dfrac{\mu_{n}}{1+ \sigma_{n} P/K}\left(1-e^{-\rho_{n} C_{n}}\right)Q,
\end{array}
\label{eq:10}
\end{equation}
where the functions $F_{J}(M,S,Q)$ represent the kinetic equations for each compartment respectively, with $J=\{M,S,Q\}$. Note that a new parameter $\epsilon$ has been introduced in the third equation. It is linked to the extent to which non-specific drugs destroy cells in the quiescent phase. It is frequently assumed that, because of their mechanism of action, CNS drugs destroy quiescent cells to the same extent as they damage cycling cells \cite{gard2}. From a logical point of view, this would be certainly so, even only by definition. However, although the mechanism of action can be clearly identified sometimes for a drug, on many occasions the death of the cell is only made effective once it reenters the cell cycle \cite{prp}. The notion of a cell kill that is totally independent of the kinetic status of the cell clearly challenges the paradigm of mitotoxicity. Even so, the fact that the proliferation rates in many chemosensitive human cancers are low clash with such paradigm, suggesting to some extent that the mechanisms of destruction might be different as those conventionally claimed \cite{perry,prp,putten}. Therefore, a value $\epsilon$ in the interval $[0,1]$ is deserved to modulate the effect of CNS drugs on resting cells.

\subsection{Pharmacokinetics and protocols of chemotherapy}
It remains to be introduced the pharmacokinetics of the present model, which determines the dynamics of the concentration of each drug $C(t)$ at the tumor site. As in previous works on chemotherapy \cite{validbul,mixed,pinho}, we assume a one-compartment model and first order pharmacokinetics, which in some situations can be considered as a good approximation to the clinical practice \cite{pks}. Hence, the differential equation governing the concentration of the drug simply reads
\begin{equation}
\dfrac{d C}{d t} = I(t) - k_{e}C,
\label{eq:11}
\end{equation}
where $I(t)$ is the function representing the rate of drug flow into the body (\emph{i. e.} the instantaneous dose-intensity) while $k_{e}$ is the rate of elimination of the drug from the bloodstream. The half-life can be computed from this constant rate by taking $\ln 2/k_{e}$.

Concerning the drug delivery, we assume that every drug is administered intravenously at constant speed during a time $\tau_{a}$ in cycles of period $T$. Therefore, a typical protocol of chemotherapy can be schematically represented as in Fig.~\ref{fig:3}. If the total drug dose given in a course of chemotherapy is $D$, then the instantaneous dose-intensity $I(t)$ is mathematically expressed as the square function
\begin{equation}
I(t)=\left\{
\begin{aligned}
\dfrac{D}{\tau_{a}} &~~\text{for}~~t(\text{mod}~T) \in [0,\tau_{a}) \\
0~&~~\text{for}~~t(\text{mod}~T) \in [\tau_{a},T)
\end{aligned}
\right..
\label{eq:12}
\end{equation}

\begin{figure}
\centering
\includegraphics[width=0.7\linewidth,height=0.3\linewidth]{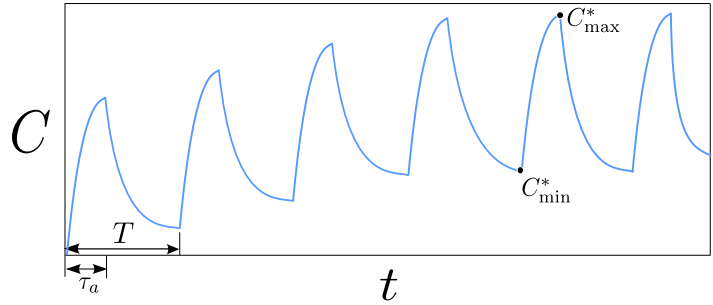}
\caption{\textbf{The drug concentration}. The time series of the drug concentration for six cycles of chemotherapy administered every $T$ weeks. Each course of chemotherapy consists of a continuous infusion of drug that lasts a time $\tau_{a}$. Hence the rate of infusion is mathematically represented by a square function of height $D/\tau_{a}$. The constant rate of elimination of the drug from the bloodstream is $k_{e}$. Asymptotically, a steady oscillation between two values of the concentration $C_{\min}^{*}$ and $C_{\max}^{*}$ is attained.}
\label{fig:4}
\end{figure}

The solution to equation \eqref{eq:11} with a drug input given by equation \eqref{eq:12}, when a number of $N+1$ chemotherapeutic cycles have been delivered, yields
\begin{equation}
C(t)=\left\{
\begin{aligned}
\dfrac{D}{k_{e} \tau_{a}}(1-e^{-k_{e} t})+a_{N} (t)~~~~~~&~~\text{for}~~t(\text{mod}~T) \in [0,\tau_{a}) \\
\dfrac{D}{k_{e} \tau_{a}}(e^{k_{e} \tau_{a}}-1)e^{-k_{e} t}+a_{N} (t)&~~\text{for}~~t(\text{mod}~T) \in [\tau_{a},T)
\end{aligned}
\right.,
\label{eq:13}
\end{equation}
where $a_{N}(t)$ represents the decaying accumulated drug concentration at the $N+1$-th cycle of chemotherapy, which equals zero for the first cycle $(a_{0}(t)=0)$ and it is equal to
\begin{equation}
a_{N} (t)=\dfrac{D}{k_{e} \tau_{a}}(e^{k_{e} \tau_{a}}-1)e^{-k_{e} t} \sum_{n=1}^{N}e^{-n k_{e} T}
\label{eq:14}
\end{equation}
for the subsequent cycles. As shown in Fig.~\ref{fig:4}, when the number of cycles tends to infinity ($N \rightarrow \infty$), the concentration reaches a steady state periodic oscillation $C^{*} (t)$ between a minimum value
\begin{equation}
C^{*}_{\min}=\dfrac{D}{k_{e} \tau_{a}}\dfrac{e^{k_{e} \tau_{a}}-1}{e^{k_{e} T}-1}
\label{eq:15}
\end{equation}
and a maximum value $C_{\max}^{*}=C_{\min}^{*}e^{k_{e}(T-\tau_{a})}$. Since the period between cycles of chemotherapy commonly spans a few weeks, whereas the drug is eliminated rather quickly in comparison (around a day), the accumulation of drug is not usually very important, unless $\tau_{a}$ is comparable to $T$.

\section{Results} \label{sec:R}
The most important part of the results presented here concerns the randomized computational study along which we compare several types of protocols in the six-dimensional parameter space spanned by the parameters $\mu_{s},\mu_{n},\rho_{s},\rho_{n},\sigma_{s}$ and $\sigma_{n}$. However, to acquire some intuition on the differences between cycle specific and cycle non-specific drugs, we thoroughly investigate the role of these drugs separately in the first place. These prior computations will enable a discussion on the role of the parameter $\epsilon$, which describes the relative effect of CNS drugs on resting cells, compared to their cycling counterparts. Moreover, they will permit us to uncover the evolution of the proliferative index along the treatment. As shown promptly, the proliferative fraction acts as a limiting factor to dose-densification when the quiescent compartment is barely chemosensitive.
\begin{figure}
\centering
\includegraphics[width=0.9\linewidth,height=0.35\linewidth]{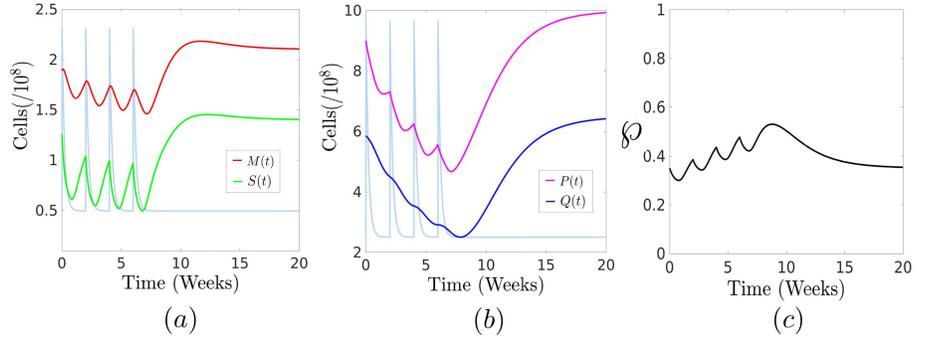}
\caption{\textbf{Administration of a single CS drug}. The evolution of the three cell populations when four cycles of $D_{s}=90~\text{mg}$ of a s-specific drug are delivered every two weeks intravenously ($\tau=1~\text{min}$). The drug dose is plotted in the background for clarity, disregarding its specific values. (a) The evolution of dividing cells in the mitotic (red) and the synthetic (green) compartments. (b) The quiescent population (blue) reduces continuously since these cells enter the cell cycle as the replicating population is reduced. The evolution of the whole tumor (magenta) is also shown for completeness. (c) The proliferative fraction increases modestly in average, because the quiescent compartment is continuously feeding the dividing one.}
\label{fig:5}
\end{figure} 

As far as the kinetic parameters of the model are concerned, we give their values in this preamble. Such values shall be used for all the simulations, unless otherwise specified. We consider a tumor whose maximum rate of growth is $r=0.9~\text{weeks}^{-1}$ and with a value of the carrying capacity of $K=10^9$ cells, which are derived from experimental values in references \cite{validpilis,validbul}. An extremely fast growing tumor would be one in which all cells were ceaselessly dividing through mitosis in the absence of apoptosis. Since the cell cycle of a human cell lasts approximately one day, this corresponds to an exponentially growing tumor with a constant rate value of $r=4.85~\text{weeks}^{-1}$. Therefore, we are considering a quite aggressive tumor for which the rate of apoptosis is rather low. The carrying capacity corresponds to a detectable premetastatic tumor mass of approximately one gram. We recall that limiting values of deadly tumor masses round numbers of $K=10^{12}$ cells \cite{perry}. The original size of the tumor is considered to be $P_{0}=9 \times 10^8$ cells, divided among the three compartments as given by their equilibrium distribution, represented by equation $\eqref{eq:6}$. We take a typical value of $k_{e}=4.85~\text{weeks}^{-1}$ for the CNS drug and $k_{e}=9.55~\text{weeks}^{-1}$ for the CS one. These two values correspond to a half-life of approximately one day and less than a half a day, respectively \cite{ncc}. The proliferative index is considered to be $\wp^{*}=0.35$, which is a reasonable intermediate value within the range appearing in previous works based on empirical data \cite{gard2}. The value of $p$ is chosen to be $0.35~\text{weeks}^{-1}$, which is exactly the same value assumed in the last cited work, and which is also based on experimental evidence \cite{killman,haemo,papaya}. This value settles the constant rate of necrosis to approximately $n=0.3~\text{weeks}^{-1}$ by means of equation~\eqref{eq:8}. Again, this amount is also of the order of other values appearing in the literature \cite{gard2}, which results in a necrotic fraction compatible with the experimental data gathered from \emph{in vivo} experiments \cite{necro}. The value of $q$ is derived from equation~\eqref{eq:5} after an elementary computation, and results to be $q =1.95 \times 10^{-9}~\text{weeks}^{-1}\text{cell}^{-1}$. To conclude, since the time spent in the s-phase of the cycle of a typical human cell is two fifths of the whole cycle, we have $f=0.4$. As we argued before $u=r/f$ and $v=r/(1-f)$, thus we obtain that $u=2.25~\text{weeks}^{-1}$ and $v=1.5~\text{weeks}^{-1}$.
\begin{figure}
\centering
\includegraphics[width=0.9\linewidth,height=0.65\linewidth]{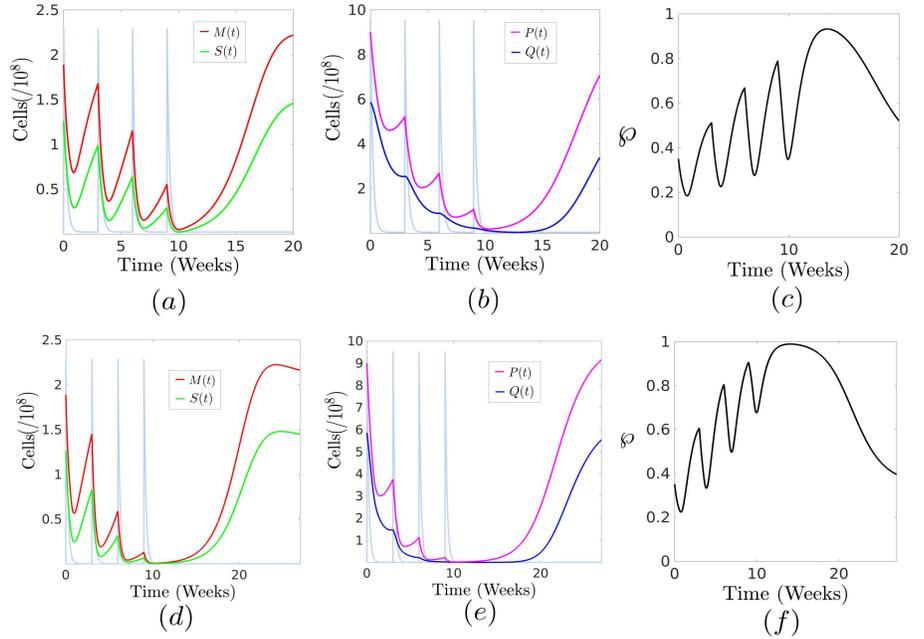}
\caption{\textbf{Administration of a single CNS drug}. The evolution of the three cell populations when four cycles of $D_{n}=60~\text{mg}$ of a non-specific drug are delivered every three weeks intravenously ($\tau_{n}=1~\text{min}$) assuming that $\epsilon=0$ in the first case, which means that quiescent cells are not affected by CNS drugs. Then a moderate destruction with $\epsilon=0.2$ is considered. (a) The evolution of cells in the mitotic (red) and the synthetic (green) compartments. (b) The quiescent population (blue) reduces continuously again. The evolution of the whole tumor (magenta) is plotted as well. (c) The proliferative fraction increases now very steeply, since the number of dividing cells increases as the quiescent cells flow from the proliferating compartment. (d-f) The same corresponding figures for the case $\epsilon=0.2$, where quiescent cells are being destroyed by the drug as well. As expected, the proliferative index reaches even higher values than for the previous case.}
\label{fig:6}
\end{figure}

\subsection{Cycle specific and cycle non-specific drugs separately} \label{sec:hcd}

We begin by studying the role of the administration of a CS drug alone. If desired, we consider $\mu_{n}=0.0~\text{weeks}^{-1}$ in the model equations \eqref{eq:10}. The value of the sensitivity is set to $\rho_{s}=1.0~\text{mg}^{-1}$, which is within values appearing in previous works for cells sensitive to methotrexate \cite{gard}, an antifolate drug that acts during the s-phase of the cycle. For reasons explained somewhere else \cite{ddpc,ncc}, the maximum fractional cell kill is set to $\mu_{s}=4.0~\text{weeks}^{-1}$. As shown in previous works \cite{ncc}, this value corresponds to a moderate value of a drug that is capable of tumor shrinking. Finally, the parameter $\sigma_{s}=1.0$ is set, corresponding to a value for which the Norton-Simon hypothesis applies rather weakly \cite{ncc}. It means that the rate of kill is reduced to half of its maximum value as the tumor gets close enough to the carrying capacity. A dose of a CS drug $D_{s}=90~\text{mg}$ is administered intravenously ($\tau_{s}=1~\text{min}$) every two weeks ($T_{s}=2~\text{weeks}$). The values of the dose and the drug periodicity are within values frequently appearing in standard protocols of chemotherapy\footnote{Information about standard protocols of chemotherapy has been drawn from $<http://www.bccancer.bc.ca/health-professionals/clinical-resources/chemotherapy-protocols>$}. For example, the CS drug methotrexate can be administered at high doses of $D={50~\text{mg}~\text{m}^{-2}}$ for an adult male \cite{hdm}. Since the average surface body area of an adult male is around $1.8~\text{m}^{2}$, a value of $D_{s}=90~\text{mg}$ can be estimated. Nevertheless, we recall that a great variability in the parameters $\mu$, $\rho$ and $\sigma$ is expected in practice \cite{gard}. For this reason, we design the randomized numerical trial presented in the next section.
\begin{figure}
\centering
\includegraphics[width=0.9\linewidth,height=0.35\linewidth]{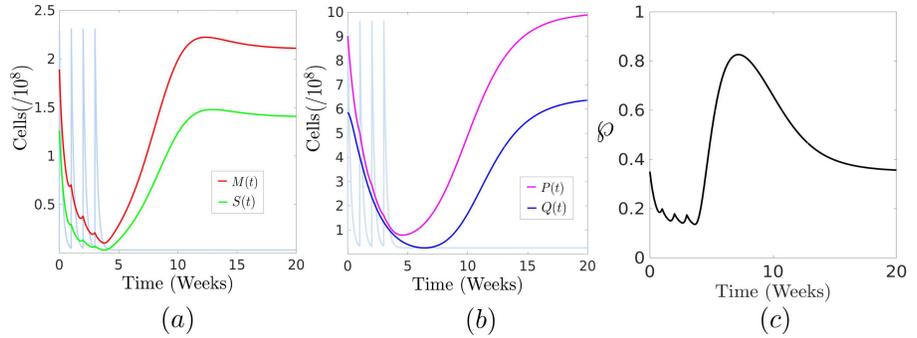}
\caption{\textbf{A limitation to dose-density}. The evolution of the three cell populations when four cycles of a non-specific drug at a dose $D_{n}=60~\text{mg}$ are delivered every week intravenously ($\tau_{n}=1~\text{min}$), assuming that $\epsilon=0$. (a) The evolution of cells in the mitotic (red) and the (green) compartments. (b) The quiescent population (blue) reduces continuously, but now the replicating cells are not allowed to regrow between cycles, leading to a very small proliferating fraction. The evolution of the whole tumor (magenta) is plotted as well. (c) The time series of the proliferative fraction $\wp$, which is kept at a very small value close to $0.2$ due to the higher density of the protocol.}
\label{fig:7}
\end{figure}

As shown in Figs.~\ref{fig:5}(a) and (b), when a CS drug destroys tumor cells in its specific compartment, this cell population is naturally reduced. As cells in the s-phase are destroyed, less cells flow from the synthesis phase to the mitotic one, so that $M(t)$ mimics $S(t)$, although the variations are less pronounced, more smooth and present a delay of a couple of days. Once the drug is eliminated from the organism, the tumor cells in the mitotic pocket regrow during the remaining part of the cycle. As the dividing tumor cells are wiped out, the resting cells abandon their state of quiescence and reenter the cell cycle. This effect produces a continuous reduction of the quiescent compartment. As can be seen in Fig.~\ref{fig:5}(c), the regrowth of tumor cells between cycles of chemotherapy and the draining of resting cells reinforce each other contributing to an average increase of the proliferative index. However, this increase and the periodic oscillations around it are very modest for CS drugs.
\begin{figure}
\centering
\includegraphics[width=0.55\linewidth,height=0.4\linewidth]{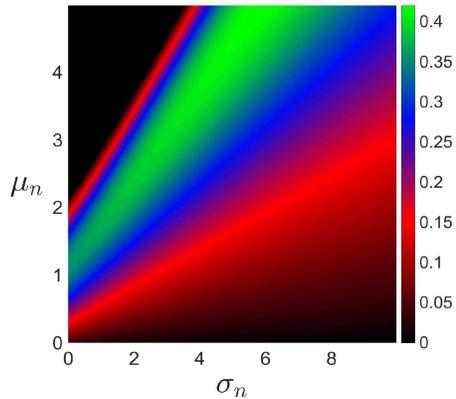}
\caption{\textbf{Parameter space}. The survival fraction at the nadir of the treatment $s_{f}=P(t_{\text{nad}})/P(0)$ is computed for a square grid of values in the parameter space $(\sigma_{n},\mu_{n})$ of the CNS drug. The values of the parameter $\mu_{n}$ are given relative to $r$. A periodic protocol of $D_{n}=60~\text{mg}$ administered intravenously with $T_{n}=3~\text{weeks}$ is computed. Then the same values are computed for a denser protocol with $T_{n}=1~\text{week}$, and the $\ln(s_{f}/\tilde{s}_{f})$ is shown, where the tilde stands for the denser protocol. The log-ratio corresponds to the colorbar, where the black color has been assigned to values smaller or equal to zero. We see that for high values of $\mu_{n}$ and small values of $\sigma_{n}$ dose-density fails to reduce the survival fraction at the nadir and might be even counterproductive.}
\label{fig:8}
\end{figure}

Now we examine the effect of administering CNS drugs on the model. For this purpose, we consider a dose of drug $D_{n}=60~\text{mg}$ administered intravenously ($\tau_{s}=1~\text{min}$) every three weeks ($T_{n}=3~\text{weeks}$), which are typical for the drug doxorubicine, used in chemotherapy for locally advanced breast cancer. The parameter values used in these simulations are $\mu_{n}=4.0~\text{weeks}^{-1}$ for the maximum fractional cell kill, $\rho=1.0~\text{mg}^{-1}$ for the sensitivity and $\sigma=1.0$ for the parameter simulating the Norton-Simon effect. In these simulations we assume that CNS drugs do not affect the quiescent compartment, thus $\epsilon=0$. As shown in Fig.~\ref{fig:6}(a) and (b), CNS drugs destroy tumor cells in the whole dividing compartment, producing a greater tumor cell decay. However, the regrowth between cycles is also higher now, and since the quiescent cells are resuming their cell cycle, these effects synergize to produce oscillations that grow in amplitude shooting up the proliferative index. In Figs.~\ref{fig:6}(d) and (e) the same results are investigated, by considering a certain destruction of quiescent cells by the CNS drug, as is commonly accepted \cite{gard2,prp}. However, following a principle of prudence, we consider a small value $\epsilon=0.2$, which is respectful with the  paradigm of mitotoxicity as well \cite{perry}. This phenomenon produces an even stronger boost of the proliferative index. Now its oscillations are smaller, because the transfer of cells from the quiescent compartment reduces slightly. We note that a possible and simplified way of modeling the most strict version of the Norton-Simon hypothesis, would be to consider the parameters $\epsilon$ and $\sigma$ both equal to zero.

These results clearly suggest that, contrary to the philosophy of dose-density, allowing some time between cycles has the benefit of maintaining a high proliferative fraction. In turn, it is then expected that if we push the dose-density too much, protocols might become less effective, unless quiescent cells are as chemosensitive as cancer cells are. This idea is corroborated when we compute the same protocol in Figs.~\ref{fig:7}(a) and (b) with the same drug but at an increased frequency of one week. The resulting plot for the proliferative fraction (see Fig.~\ref{fig:7}(c)) indicates that some time is required for the quiescent cells to reenter the cell cycle and regrow, posing a limit on the frequency reduction craved by dose-dense protocols. Furthermore, and as a completely novel result, we show in Fig.~\ref{fig:8} that when comparing the survival fraction at the nadir $s_{f}=P(t_{\text{nad}})/P(0)$  of two treatments for $\epsilon=0$, one being three times more dense that the other, there exist parameter values for which dose-density is not recommendable. We consider this statement an important revelation of the present model compared to previous investigations on this topic, where the dose-dense principle was unequivocally advocated \cite{ddpc,ncc}.
\begin{figure}
\centering
\includegraphics[width=1.0\linewidth,height=0.7\linewidth]{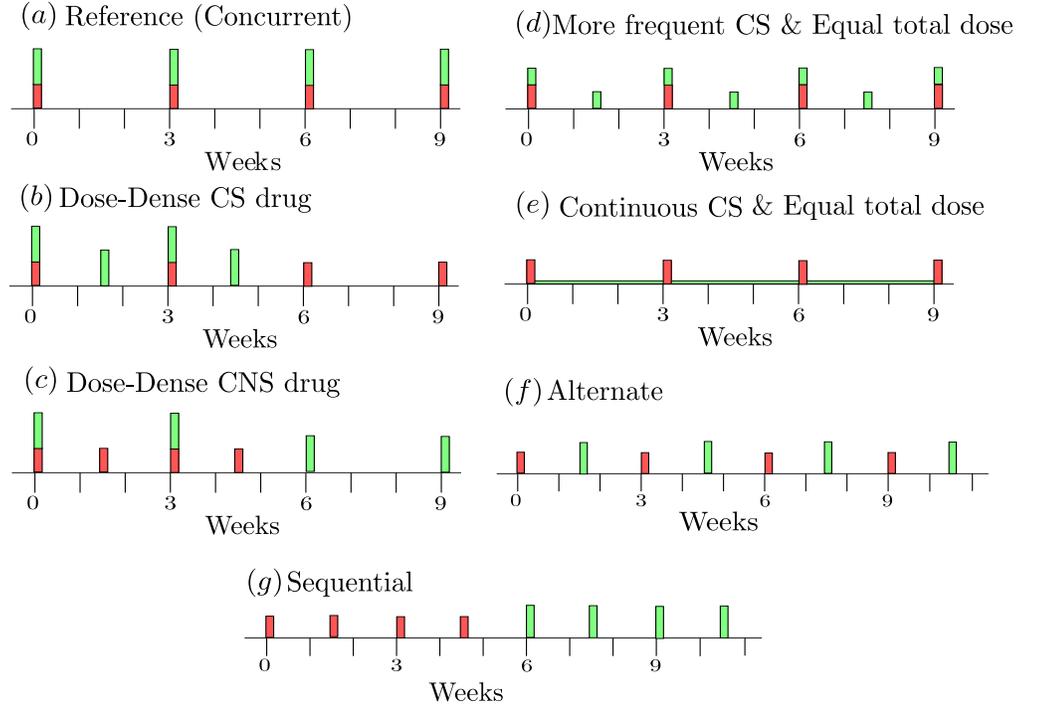}
\caption{\textbf{Randomized numerical trial}. A total number of seven chemotherapeutic protocols are compared among them. Unless otherwise specified, all the drugs are administered intravenously ($\tau=1~\text{min}$). (a) The reference protocol, for which a CS and a CNS drug are given concurrently every three weeks at doses $D_{s}=90~\text{mg}$ and $D_{n}=60~\text{mg}$, respectively. (b) A protocol that is twice denser than the reference in the CS drug. (c) A protocol that is twice denser than the reference in the CNS drug. (d) A protocol with a higher frequency but $D_{s}=45~\text{mg}$, which equals to the same total dose. (e) A protocol with the same total dose for the CS drug as the reference ($360~\text{mg}$), but administered continuously. (f) A protocol for which the two drugs are given alternately every one and a half weeks. (g) A sequential protocol, where one of the drugs (CNS) is given in the first place every one and a half weeks, while the second is given right after (CS), with identical frequency.}
\label{fig:9}
\end{figure}

\subsection{Randomized computational trial}\label{sec:nsi}
Now we accomplish a randomized comparative study between several types of chemotherapeutic protocols. Inspired by clinical trials, we represent a vast number of different virtual patients (so to speak) by letting the parameters $\mu_{s}$, $\mu_{n}$ fluctuate in the range, $[0.1,4.5]$ with a uniform distribution. The parameters $\sigma_{s}$ and $\sigma_{n}$ are selected from the interval $[0.1,5.0]$ following the same recipe. Finally, the values of $\rho_{s}$, $\rho_{n}$ are picked at random in the range $[0.1,1.0]$ and $\epsilon$ is considered to be equal to $0.2$. This set of values is enough to model both chemosensitive and resistant cell clones \cite{ncc}, a considerable heterogeneity of drug responses to the drugs and a considerable degree of mitotoxicity. As far as we have investigated, doubling the size of the parameter intervals does not produce a substantial variation of our results. We use a total count of $400.000$ parameter values for each protocol, which has proven to be sufficient to give a convergent probability distribution at different scales and allow fast computations at the same time. Finally, for each point in the six-dimensional parameter space and every protocol, we compute the survival fraction at the nadir $s_{f}=P(t_{\text{nad}})/P(0)$, using the same initial conditions as in the simulations of the previous section. This effectiveness criterion has been adopted in previous works \cite{gard2,ncc}. Consequently, we recall that, since in the present study we are not explicitly modeling toxicity, the benefits or the effectiveness of a protocol should not be read from a clinical point of view, but from a theoretical one instead.
\begin{figure}
\centering
\includegraphics[width=0.7\linewidth,height=0.3\linewidth]{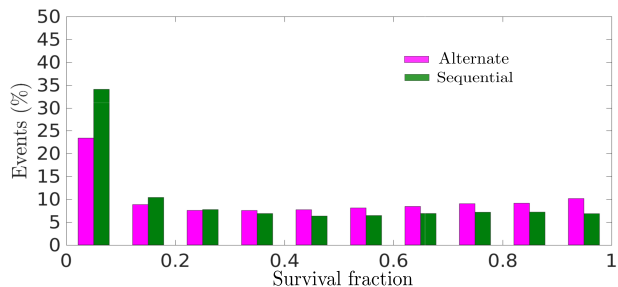}
\caption{\textbf{Sequential or alternating?} A numerical trial comprising 400.000 random and uniformly chosen events in the six-dimensional parameter space $(\mu_{s},\sigma_{s}, \rho_{s},\mu_{n},\sigma_{n},\rho_{n})$ is carried out. The parameters $\mu$ and $\sigma$ take values in the intervals $[0.1,4.5]$ and $[0.1,5.0]$ respectively, while the parameters $\rho$ are selected in the range $[0.1,1.0]$. The number of events with a particular survival fraction at the nadir of the treatment are shown for the sequential (green) and the alternating (magenta) regimes, showing the greater benefits of the former.}
\label{fig:10}
\end{figure}

We now briefly discuss the design of the protocols. As seen in Fig.~\ref{fig:9} seven different protocols are computed. A first reference protocol for which the two drugs (CS and CNS) are delivered concurrently every three weeks at doses appearing in the previous sections. The duration of the infusions also remains unchanged ($\tau=1~\text{min}$) with respect to the previous simulations. Then, two more protocols are computed to test the benefits of dose-densification in relation to the specificity of the drug. The former protocol is twice denser in the CS drug, while the latter is more tight for the CNS drug. The frequency is doubled in order to test protocols that are limiting in the time scale frequently used in the clinical practice (order of weeks). Now, the two next protocols are conceived to test an alternative way of avoiding tumor growth between cycles, compared to dose-densification. For this purpose, we use the CS-drug administered at constant total doses, and increase progressively the frequency of drug administration. Therefore, the fourth protocol for which the CS drugs is given at half the original dose  ($D_{s}=45~\text{mg}$), but delivered every one and a half weeks. Then, the same total dose of $D_{s}=360~\text{mg}$ is delivered continuously. In this last case the drug concentration is very small to destroy the tumor, but enough to halt its regrowth between cycles of the CNS drug. Finally we introduce two additional protocols to compare the relative benefits of sequential against alternating protocols. To some extent, this evaluation allows us to ascertain the good performance of our model, since they are similar to real life trials carried out in the recent past \cite{dd1,norton}.

We begin by discussing the differences between the sequential and the alternating arms. As can be seen in Fig.~\ref{fig:10}, the model echoes the conclusions discussed in previous studies \cite{norton}. The number of events with low survival fractions are much smaller for the alternating protocol than for the sequential arm. In particular, we find that the latter is superior by more than a $10\%$ of events with one-log kill. This means that sequential administration is a better strategy, insofar as it is more dense in each drug. However, and as discussed in \cite{nsh}, this is especially true in this model because one of the drugs of the protocol is more effective than the other. Clearly put, our model does not tell the difference between the particular biochemical mechanisms of each drug, and therefore sequential and alternating protocols can not be distinguished by any means, as long as such drugs cause similar tumor cell destruction.

Concerning dose-dense protocols, the results appearing in Fig.~\ref{fig:11} leave no doubt about the benefits of dose-densification for CNS drugs, where we obtain around a $13\%$ of events with one more log kills compared to the reference protocol. Unexpectedly, an increase in the frequency of drug administration does not seem to be a recommendable strategy if the CS drug is considered, since it barely augments the number of events with small survival fractions. In contrast, the number of events with survival fractions closer to one is noticeably higher. Probably, the reason behind this phenomenon is that when the cells in the synthesis phase are erased, it is useless to keep on trying to destroy this compartment, since it is already empty. Thus, this effect would be the dose-dense analog to the variation of the dose-response curves commonly observed for CS drugs, which limits the benefits of their dose intensification.
\begin{figure}
\centering
\includegraphics[width=0.7\linewidth,height=0.3\linewidth]{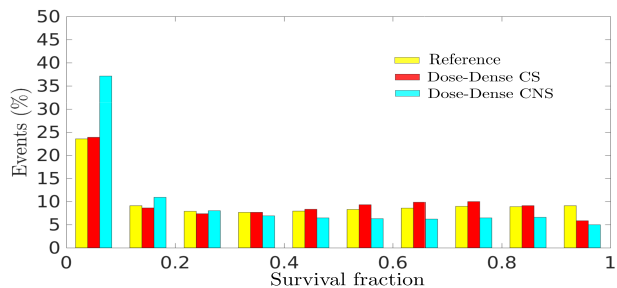}
\caption{\textbf{Dose-densification}. A numerical trial comprising 400.000 random and uniformly chosen events in the six-dimensional parameter space $(\mu_{s},\sigma_{s}, \rho_{s},\mu_{n},\sigma_{n},\rho_{n})$ is carried out. The parameters $\mu$ and $\sigma$ take values in the intervals $[0.1,4.5]$ and $[0.1,5.0]$ respectively, while the parameters $\rho$ are selected in the range $[0.1,1.0]$. Three protocols are tested, being the first the concurrent reference protocol, the second the more dose-dense in the CS drug, while the third being denser in the CNS drug. The number of events with a particular survival fraction at the nadir of the treatment for the reference (yellow), the CS-dense (red) and the CNS-dense (cyan) are computed, showing the outstanding benefits of the later.}
\label{fig:11}
\end{figure}
Regarding the possible benefits of protocols where the CS drug is given more frequently but at a constant equal total dose, we have found the following results. When the frequency is doubled and the dose per cycle is reduced to half of its control value, we see a modest increase in the protocol. However, when the CS drug is delivered continuously, despite being at quite low doses, we obtain a tremendous increase in overall tumor reduction. More specifically, a $ 20\%$ increase in cases bellow one-log kill is recorded. As we are about to show by going to smaller scales of the survival fraction, the impact is even more striking for this protocol. Accordingly, and as a colophon to this randomized numerical trial, we compare the relative benefits obtained for the third and the fifth protocols, which have been the two most outstanding settings.

\begin{figure}
\centering
\includegraphics[width=0.95\linewidth,height=0.55\linewidth]{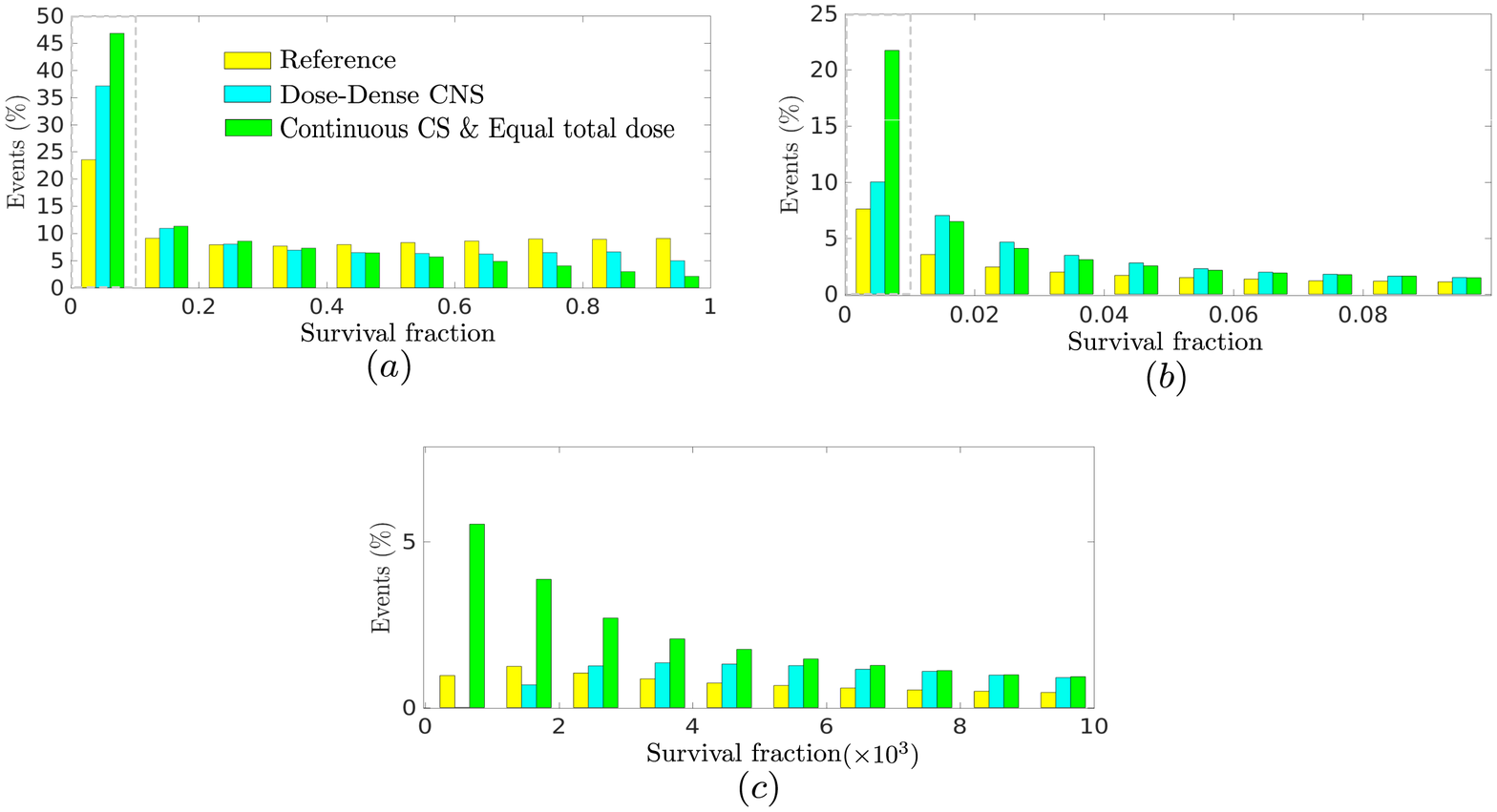}
\caption{\textbf{Comparing between the best strategies}. The randomized numerical trials previously discussed are shown for the reference protocol, the CNS-dose-dense delivered twice more frequently and the continuously delivered protocol at low doses, as a third arm. (a) The number of events with a particular survival fraction at the nadir for the reference (yellow), the more frequent in the CNS drug (cyan) and the continuous protocol in the CS drug (green), clearly depicting the outstanding benefits of the later. (b) A zoom of the region appearing in the dashed box in the previous figure. More than $10\%$ of the cases reach one-log of cell kill when regrowth is avoided by using the protocol that is continuous in the CS drug, instead of a dose-dense protocol in the CNS drug. (c) Another zoom of the region appearing in the dashed box of the second figure is shown. As can be seen, more than $5\%$ of the cases reach three-log kills when regrowth is avoided by using the protocol that is continuous in the CS drug.}
\label{fig:13}
\end{figure}

These results are contrasted in Figs.~\ref{fig:13}(a). Not only the continuous CS protocol is about $6 \%$ more destructive under one-log kill than the CNS-dense protocol. When the results are plotted by zooming inside the region of survival fractions between $0$ and $0.1$, we find that more than $10 \%$ of the cases for the CS-continuous protocol are under two-log kills. Moreover, by pushing further our scrutiny, we find that a non negligible $6 \%$ of the cases lie under three-log kills.  We believe that these results are very relevant from a theoretical point of view and, as discussed in the next section, have the virtue of suggesting a possible change of paradigm concerning the future of dose-dense chemotherapy.

\begin{figure}
\centering
\includegraphics[width=0.7\linewidth,height=0.3\linewidth]{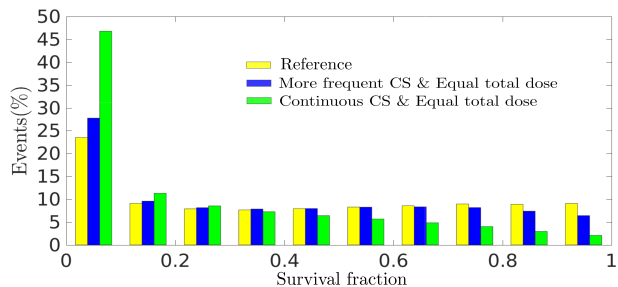}
\caption{\textbf{Continuation at same total dose}. A numerical trial comprising 400.000 random and uniformly chosen events in the six-dimensional parameter space $(\mu_{s},\sigma_{s},\rho_{s}, \mu_{n},\sigma_{n},\rho_{n})$ is carried out. The parameters $\mu$ and $\sigma$ take values in the intervals $[0.1,4.5]$ and $[0.1,5.0]$ respectively, while the parameters $\rho$ are selected in the range $[0.1,1.0]$. The concurrent protocol is shown as a reference, together with a protocol that is more dense but at an smaller dose per cycle, and the protocol with continuously delivered CS drug as the third contending protocol. The number of events with a particular survival fraction at the nadir for the reference (yellow), the more frequent in the CS drug (blue) and the continuous protocol in the CS drug (green), clearly depicting the outstanding benefits of the later.}
\label{fig:12}
\end{figure}

\section{Discussion and Conclusions}\label{sec:con}

In the present paper we have introduced a quasispecies model of tumor growth for cancer chemotherapy that allows to represent tumor masses formed by cells with different kinetic status. Even though the dimensionality has been kept low in this initial study, the extension of the model to reproduce heterogeneous tumors is straightforward and can serve to explore the evolution of resistance in maintenance or palliative chemotherapy, in addition to adjuvant therapy. Moreover, the incorporation of other players to the dynamical scenario, as for example the cell-mediated immune response or other types of cells from mesenchymal origin, can enable a more accurate modeling of mixed therapies \cite{mixed} and permit the simulation of toxicity as well.

The kinetic variability of our mathematical ODE model has allowed us to unveil limitations on the principle of dose-densification that were unobserved in previous works \cite{ddpc,ncc}. These conclusions have been drawn on two different fronts. Firstly, our study sheds light into the role of specific drugs in dose-dense chemotherapy. As strongly suggested by our simulations, dose-densification of cycle-specific drugs is envisaged to be counterproductive. On the other hand, even though CNS drugs and dose-densification are expected to be generally beneficial, too dense protocols might not allow the quiescent compartment to flow into the replicating one, rendering dose-density less effective than \emph{a priori} foreseen. In fact, we wonder if this effect is behind the failure of dose-density that  has been reported recently in trials of dense cancer chemotherapy \cite{notwo}. Evidently, this effect arises as long as the destruction in the quiescent compartment is not too high compared to the mitotic one. Otherwise, as $\epsilon$ is made close to $1.0$, dose-density for CNS drugs would be unconditionally recommended.

Secondly, our study suggests that other possible strategies can be devised to arrest the regrowth of the tumor between cycles. In the present work, we have proposed that the use of CS drugs at low doses administered continuously in addition to CNS drugs periodically delivered can be a much better schedule than dose-densification in the CNS drug alone. The authors are aware that continuous protocols can lead to tachyphilaxis and might also be too toxic \cite{dan}, even for CS drugs. Therefore, it is not our desire to suggest this specific strategy as a viable optimal choice, but rather to set off a debate concerning alternatives to dose-densification, capable of producing the same effect in a more forceful and safe way. 

As an example, in previous works \cite{ddpc} we reasoned that the use of targeted therapies using cytostatic agents to arrest the tumor revival between cycles might serve to complement cytotoxic chemotherapeutic drugs. Certainly, this way of tackling the problem presents other difficulties \cite{ddpc}, but we note that a change in the point of view of dose-dense chemotherapy is entailed by this perspective. In lieu of trying to achieve maximum fractional cell kill by simply reducing the time between cycles, we would rather put the tumors to sleep during the meantime. In this way, stasis and destruction would cooperate to produce an enhanced overall destruction of the tumor. Among other potential benefits, this approach could allow to extend recovery times between cycles of chemotherapy. We should not forget that in order to damp the increased toxicity of dose-dense protocols, growth factors are currently delivered to the patients.

Thus, to conclude, we retake once last Paul Ehrlich's old adage and suggest that the periodically thrown cannonballs of cytotoxic chemotherapy might be complemented with the magic bullets of some cytostatic targeted therapy administer in between \cite{mbul}. Then, the following hesitation immediately arises. In order to produce the aforementioned effect without halting the rebound of the collaterally damaged healthy tissues, can cytostatic drugs be readily designed to impact more selectively the tumors than cytotoxic ones?

\section*{Aknowledgments}  \label{sec:aknow}
This work has been supported by the Spanish Ministry of Economy and Competitiveness and by the Spanish State Research Agency (AEI) and the European Regional Development Fund (FEDER) under Project No. FIS2016-76883-P. K.C.I. acknowledges FAPESP (2015/07311-7 and 2018/03211-6).


\section*{Additional information}
\textbf{Competing financial interests}: The authors declare to have no competing financial interests.

\end{document}